\documentstyle[aps,preprint]{revtex}
\input epsf
\global\firstfigfalse  % kludge to bypass foolishness of revtex to place

\tighten

\begin{document}
\pagestyle{empty}

\preprint{
\begin{minipage}[t]{3in}
%\begin{flushright} IAS--TH--00/18,  NSF-ITP-00-16 \\
%hep-th/0003136 \\
%March 2000
%\end{flushright}
\end{minipage}
}

\title{Statefinder diagnostic for coupled quintessence}

\author{Xin Zhang \\\bigskip}
\address{Institute of High Energy Physics, Chinese Academy of Sciences\\
P.O.Box 918(4), 100039 Beijing, China\\
\smallskip{\tt zhangxin@mail.ihep.ac.cn}}

\maketitle

\begin{abstract}

The problem of the cosmic coincidence is a longstanding puzzle.
This conundrum may be solved by introducing a coupling between the
two dark sectors. In this Letter, we study two cases of the
coupled quintessence scenario. $(a)$ Assume that the mass of dark
matter particles depends exponentially on the scalar field
associated to dark energy and meanwhile the scalar field evolves
in an exponential potential; $(b)$ Assume that the mass of dark
matter particles depends on a power law function of the scalar
field and meanwhile the scalar field evolves in a power law
potential. Since the dynamics of this system is dominated by an
attractor solution, the mass of dark matter particles is forced to
change with time as to ensure that the ratio between the energy
densities of dark matter and dark energy becomes a constant at
late times, and one thus solve the cosmic coincidence problem
naturally. We perform a statefinder diagnostic to both cases of
this coupled quintessence scenario. It is shown that the evolving
trajectory of this scenario in the $s-r$ diagram is quite
different from those of other dark energy models.

\vskip .4cm

%\noindent PACS: 04.70.Dy;02.20.Uw;97.60.Lf

\vskip .2cm

%\noindent Keywords: Schwarzschild black hole; evaporation;
%$q$-deformation; noncommutative spacetime.

\end{abstract}

\newpage
\pagestyle{plain} \narrowtext \baselineskip=18pt

\setcounter{footnote}{0}

There are more and more evidences \cite{sn,wmap,sdss} support that
the present universe is dominated by dark sectors. Combined
analysis of cosmological observations, esp. the Wilkinson
Microwave Anisotropy Probe (WMAP) satellite experiment
\cite{wmap}, shows that dark energy (DE) occupies about $73\%$ of
the energy of our universe, and dark matter (DM) about $23\%$. The
usual baryon matter which can be described by our known particle
theory occupies only about $4\%$ of the total energy of the
universe. The accelerated expansion of the present universe is
attributed to that DE is an exotic component with negative
pressure, such as the cosmological constant \cite{cc} or a scalar
field with a proper potential (i.e. the so-called quintessence)
\cite{quin}. The cosmological constant $\Lambda$ (or vacuum
energy) has the equation of state $w=-1$. The cosmological model
that consists of a mixture of vacuum energy and cold dark matter
(CDM) is called LCDM (or $\Lambda$CDM). While the so-called QCDM
cosmology is based upon a mixture of CDM and quintessence field.
The energy density and the negative pressure are provided by the
quintessence scalar field $\phi$ slowly evolving down its
potential $V(\phi)$. The equation of state of the quintessence
$-1<w<-1/3$ is guaranteed by the slow evolution. However, as is
well known, there are two difficulties arise from all of these
scenarios, namely, the 'fine-tuning' problem and the 'cosmic
coincidence' problem. The cosmic coincidence problem
\cite{coincidence} states: Since the energy densities of DE and DM
scale so differently during the expansion of the universe, why are
they nearly equal today? To get this coincidence, it appears that
their ratio must be set to a specific, infinitesimal value in the
very early universe.

A possible solution to this cosmic coincidence problem may be
provided by introducing a coupling between quintessence DE and
CDM. This coupling is often described by the varible-mass particle
(VAMP) scenario \cite{vamp}. The VAMP scenario assumes that the
CDM particles interact with the scalar DE field resulting in a
time-dependent mass, i.e. the mass of the CDM particles evolves
according to some function of the scalar field $\phi$. In this
Letter we study two cases of this coupled quintessence scenario:
$(a)$ The quintessence scalar field $\phi$ evolves in an
exponential potential and the DM particle mass also depends
exponentially on $\phi$; $(b)$ The quintessence scalar field
$\phi$ evolves in a power law potential and the DM particle mass
also depends on a power law function of $\phi$. In both cases, the
late time behavior of the cosmological equations will give
accelerated expansion and, a constant ratio between DM energy
density $\rho_\chi$ and DE energy density $\rho_\phi$
\cite{couple,zx}. This behavior relies on the existence of an
attractor solution, which makes the effective equation of state of
DE mimic the effective equation of state of DM at late times so
that the late time cosmology insensitive to the initial conditions
for DE and DM. Therefore, the scenario containing coupled
quintessence with VAMPs solves the cosmic coincidence problem in
this sense.

In this Letter, we will first show the solution to the problem of
cosmic coincidence given by this scenario and, then we perform a
statefinder diagnostic for both cases of this coupled quintessence
model. The statefinder parameters introduced by Sahni et al.
\cite{sahni} are proven to be useful tools to characterize and
differentiate between various DE models. We show in this Letter
that the evolving trajectory of this scenario in the $s-r$ diagram
is quite different from those of other DE models.

Consider, now, the interacting DE model in which we postulate that
the DM component $\chi$ interacts with the DE field $\phi$ through
the interaction term $Q$ according to
\begin{equation}
\dot{\rho}_\chi+3H\rho_\chi=-Q~,
\end{equation}
\begin{equation}
\dot{\rho}_\phi+3H\rho_\phi(1+w_\phi)=Q~,
\end{equation} where $\rho_\chi$ and $\rho_\phi$ are energy
densities of DM and DE, respectively, dot denotes a derivative
with respect to time $t$, $H=\dot{a}/a$ represents the Hubble
parameter, in which $a(t)$ is the scale factor of the universe,
and
\begin{equation}
w_\phi={p_\phi\over\rho_\phi}={{1\over
2}\dot{\phi}^2-V(\phi)\over{1\over 2}\dot{\phi}^2+V(\phi)}~
\end{equation} is the usual parameter of equation of state for the
homogeneous scalar field $\phi$ associated to DE, where $V(\phi)$
is some potential of the quintessence field. For convenience we
can define the effective equations of state for DM and DE through
the parameters
\begin{equation}
w_\chi^{(e)}={Q\over 3H\rho_\chi}~, ~~~~~~
w_\phi^{(e)}=w_\phi-{Q\over 3H\rho_\phi}~.
\end{equation} If the effective equation of state parameters
$w_\chi^{(e)}$ and $w_\phi^{(e)}$ evolve to be an equal constant
at late times, the field system is then proven to have a stable
attractor solution. In what follows we will discuss two cases of
this coupled quintessence scenario
--- the exponential case and the power law case.

\noindent{\bf Exponential case}

Assume that the DM particle $\chi$ with mass $M$ depending
exponentially on the DE field $\phi$,
\begin{equation}
M_\chi(\phi)=M_{\ast}e^{-\lambda\phi}~,\label{m}
\end{equation}
where $\phi$ is expressed in units of the reduced Planck mass
$M_p$ ($M_p\equiv 1/\sqrt{8\pi G}=2.436\times 10^{18} {\rm GeV}$),
and $\lambda$ is a positive constant. The scalar field has an
exponential potential
\begin{equation}
V(\phi)=V_{\ast}e^{\eta\phi}~,\label{v}
\end{equation} where $\eta$ is a positive constant. As a
consequence, the interaction term $Q$ in this case can be given
\begin{equation}
Q=\lambda{\dot{\phi}}\rho_\chi~.
\end{equation} The equation of motion is then
\begin{equation}
{1\over 3}{(\rho_\chi+\rho_{\rm b}+\rho_{\rm rad}+V)\over
1-\phi'^2/6}\phi''+{1\over 2}(\rho_\chi+\rho_{\rm b}+{2\over
3}\rho_{\rm rad}+2V)\phi'={\lambda}\rho_\chi-{\eta}V~,
\end{equation} where $\rho_{\rm b}$ and $\rho_{\rm rad}$ are energy densities of
baryons and radiation, respectively, and we have assumed a
spatially flat universe. Primes denote derivatives with respect to
$u=\ln(a/a_0)=-\ln(1+z)$, in which $z$ is the red-shift, and $a_0$
represents the current scale factor. Since we are interested in
the late-time behavior, we can assume $\rho_{\rm b},~\rho_{\rm
rad}\ll \rho_\chi,~\rho_\phi$. In this limit it is easy to see
that there is a solution

\begin{equation}
\phi=\phi_0-{3\over\lambda+\eta}u~,\label{attractor1}
\end{equation} such that

\begin{equation}
\Omega_\phi\simeq
1-\Omega_\chi={3+\lambda(\lambda+\eta)\over(\lambda+\eta)^2}~,
\end{equation} and

\begin{equation}
w_\phi^{(e)}=w_\chi^{(e)}=W=-{\lambda\over\lambda+\eta}~.\label{W1}
\end{equation} This is an attractor in field space for
$\eta>(-\lambda+{\sqrt{\lambda^2+12}})/2$. When the attractor is
reached, the energy densities of DM and DE will evolve at a
constant ratio depending only on $\lambda$ and $\eta$, thus
solving the cosmic coincidence problem.

It is shown by (\ref{W1}) that $W$ is negative and may lead, if
$W<-1/3$, to an accelerated expansion of the universe. To
understand how it is possible to get both acceleration and
constant ratio between DM and DE one may look at the scaling
behavior of the energy densities on the attractor
(\ref{attractor1}),
\begin{equation}
\rho_\chi\sim e^{-\lambda\phi-3u}\sim\rho_\phi\sim
e^{\eta\phi}\sim e^{-3(1+W)u}~.
\end{equation} The scaling behavior of DM deviates from the usual
scaling way $e^{-3u}$ due to the $\phi$-dependence of the DM mass.
The interaction between DM and DE forces their effective equation
of state parameters to become an equal negative constant $W$, and
thus solving the coincidence problem and at the same time
resulting in an accelerated expansion.

%%%%%%%%%%%%%%%%%%%%%%%%%%%%%%%%%%%%%%%%%%%%%%%%%%%%%%%%%%%%%%%%$
\vskip.8cm
\begin{figure}
\begin{center}
\leavevmode \epsfbox{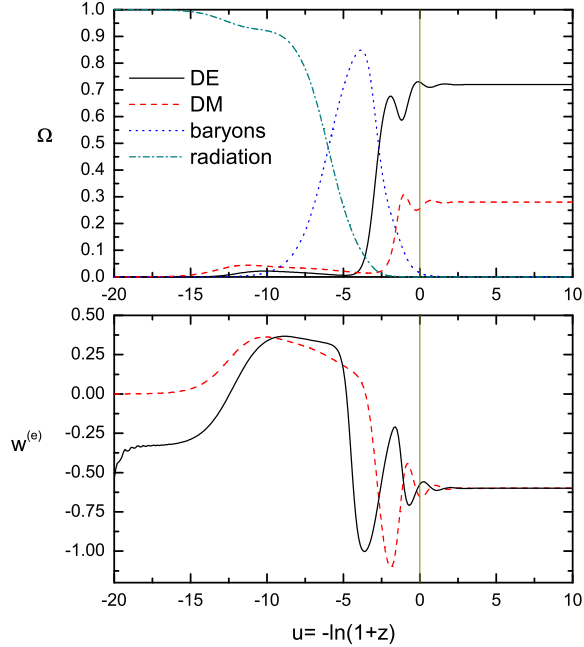} \caption[]{A typical solution to the
exponential case. The evolution of the density parameters for
different components and the effective equation of state
parameters for DE and DM. The corresponding model parameters are:
$\eta=2,~\lambda=3,~M_\ast=0.4\rho_0/n_{\chi 0}$ and
$V_\ast=0.1\rho_0$.}
\end{center}
\end{figure}
%%%%%%%%%%%%%%%%%%%%%%%%%%%%%%%%%%%%%%%%%%%%%%%%%%%%%%%%%%%%%%%%%

The time evolution of the density parameters for different
components (including also $\rho_{\rm b}$ and $\rho_{\rm rad}$)
and the effective equation of state parameters for DE and DM for a
typical solution is plotted in Fig.1. Notice that the attractor
solution is going to be reached currently in this example.

\noindent{\bf Power law case}

In this case we assume that the DM particle $\chi$ with mass $M$
depending on a power law function of the DE field $\phi$,
\begin{equation}
M_\chi(\phi)=M_{\ast}\phi^{-\alpha}~,\label{m}
\end{equation} and the scalar field has a power
law potential
\begin{equation}
V(\phi)=V_{\ast}\phi^\beta~,\label{v}
\end{equation} where $\alpha,\beta>0$. The interaction term $Q$ in this case is
then expressed as
\begin{equation}
Q=\alpha{\dot{\phi}\over\phi}\rho_\chi~.
\end{equation} The equation of motion can be given
\begin{equation}
{1\over 3}{(\rho_\chi+\rho_{\rm b}+\rho_{\rm rad}+V)\over
1-\phi'^2/6}\phi''+{1\over 2}(\rho_\chi+\rho_{\rm b}+{2\over
3}\rho_{\rm
rad}+2V)\phi'={\alpha\over\phi}\rho_\chi-{\beta\over\phi}V~,
\end{equation} and it can be proven that there is a stable
attractor solution in the field space \cite{zx,xjb}
\begin{equation}
\phi=\phi_0e^{-{3\over\alpha+\beta}u}~,\label{attractor2}
\end{equation} such that

\begin{equation}
\Omega_\phi\simeq 1-\Omega_\chi={\alpha\over\alpha+\beta}~,
\end{equation} and

\begin{equation}
w_\phi^{(e)}=w_\chi^{(e)}=W=-{\alpha\over\alpha+\beta}~.\label{W2}
\end{equation} When the attractor is
reached, the energy densities of DM and DE will evolve at a
constant ratio depending only on $\alpha$ and $\beta$, thus
solving the cosmic coincidence problem. We see that in this case
$W$ is also a negative constant and can thus lead to an
accelerated expansion of the universe. The scaling behavior of the
energy densities on the attractor (\ref{attractor2}) is exhibited
as
\begin{equation}
\rho_\chi\sim \phi^{-\alpha}e^{-3u}\sim\rho_\phi\sim
\phi^\beta\sim e^{-3(1+W)u}~.
\end{equation}

%%%%%%%%%%%%%%%%%%%%%%%%%%%%%%%%%%%%%%%%%%%%%%%%%%%%%%%%%%%%%%%%$
\vskip.8cm
\begin{figure}
\begin{center}
\leavevmode \epsfbox{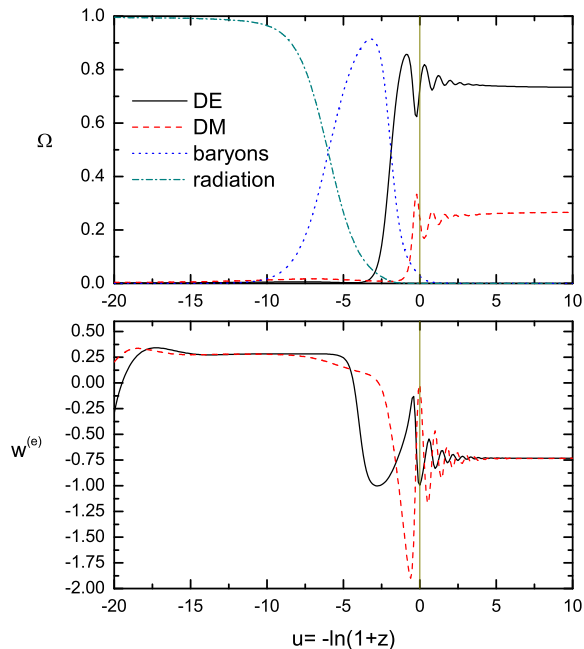} \caption[]{A typical solution to
the power law case. The evolution of the density parameters for
different components and the effective equation of state
parameters for DE and DM. The corresponding model parameters are:
$\alpha=11,~\beta=4,~M_\ast=230\rho_0/n_{\chi 0}$ and
$V_\ast=0.1\rho_0$. }
\end{center}
\end{figure}
%%%%%%%%%%%%%%%%%%%%%%%%%%%%%%%%%%%%%%%%%%%%%%%%%%%%%%%%%%%%%%%%%

In Fig.2 we plot a typical solution, including also $\rho_{\rm b}$
and $\rho_{\rm rad}$. Notice that the attractor solution is going
to be reached today in this example.

In what follows we will perform a statefinder diagnostic to this
coupled quintessence scenario. Since more and more DE models are
constructed for interpreting or describing the cosmic
acceleration, there exists the problem of discriminating between
the various contenders. In order to be able to differentiate
between those competing cosmological scenarios involving DE, a
sensitive and robust diagnostic for DE models is a must. For this
purpose a diagnostic proposal that makes use of parameter pair
$\{r,s\}$, the so-called "statefinder", was introduced by Sahni et
al. \cite{sahni}. The statefinder probes the expansion dynamics of
the universe through higher derivatives of the expansion factor
$\stackrel{...}{a}$ and is a natural companion to the deceleration
parameter which depends upon $\ddot a$. The statefinder pair
$\{r,s\}$ is defined as follows
\begin{equation}
r\equiv
\frac{\stackrel{...}{a}}{aH^3},~~~~s\equiv\frac{r-1}{3(q-1/2)}~.\label{rs}
\end{equation} The statefinder
is a 'geometrical' diagnostic in the sense that it depends upon
the expansion factor and hence upon the metric describing
space-time.

Trajectories in the $s-r$ plane corresponding to different
cosmological models exhibit qualitatively different behaviors. The
spatially flat LCDM scenario corresponds to a fixed point in the
diagram
\begin{equation}
\{s,r\}\bigg\vert_{\rm LCDM} = \{ 0,1\} ~.
\end{equation}
Departure of a given DE model from this fixed point provides a
good way of establishing the 'distance' of this model from LCDM
\cite{alam}. As demonstrated in \cite{sahni,alam,gorini,zimdahl}
the statefinder can successfully differentiate between a wide
variety of DE models including the cosmological constant,
quintessence, the Chaplygin gas, braneworld models and interacting
DE models. The interacting DE model analyzed in Ref.
\cite{zimdahl} cannot solve but only alleviate the cosmic
coincidence problem. We in this Letter will perform a diagnostic
for the coupled quintessence scenario which can provide a natural
solution to the coincidence problem and show explicitly the
difference between this scenario and other DE models.

The statefinder parameters can be expressed in terms of the total
energy density $\rho$ and the total pressure $p$ in the universe:
\begin{equation}
r=1+{9(\rho+p)\over
2\rho}{\dot{p}\over\dot{\rho}}~,~~~~s={(\rho+p)\over
p}{\dot{p}\over\dot{\rho}}~.
\end{equation} Since the total energy of the universe is
conserved, we have $\dot{\rho}=-3H(\rho+p).$ Then making use of
$\dot{\rho}_\phi=-3H(1+w_\phi^{(e)})\rho_\phi$ and
$\dot{\rho}_{\rm rad}=-4H\rho_{\rm rad}$, we can get
\begin{equation}
\dot{p}/H=p'=[w_\phi'-3w_\phi(1+w_\phi^{(e)})]\rho_\phi-{4\over
3}\rho_{\rm rad}~.
\end{equation} Hence, the statefinder parameters for the coupled
quintessence scenario can be obtained
\begin{equation}
r=1-{3\over2}[w_\phi'-3w_\phi(1+w_\phi^{(e)})]\Omega_\phi+2\Omega_{\rm
rad}~,
\end{equation}
\begin{equation}
s={-3[w_\phi'-3w_\phi(1+w_\phi^{(e)})]\Omega_\phi+4\Omega_{\rm
rad}\over 9w_\phi\Omega_\phi+3\Omega_{\rm rad}}~.
\end{equation} The deceleration parameter is also given
\begin{equation}
q={1\over2}(1+3w_\phi\Omega_\phi+\Omega_{\rm rad})~.
\end{equation}

%%%%%%%%%%%%%%%%%%%%%%%%%%%%%%%%%%%%%%%%%%%%%%%%%%%%%%%%%%%%%%%%$
\vskip.8cm
\begin{figure}
\begin{center}
\leavevmode \epsfbox{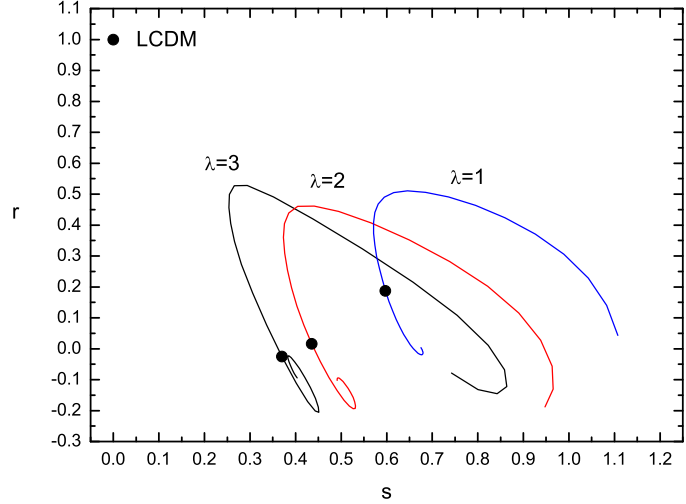} \caption[]{The $s-r$ diagram of the
exponential case: evolution trajectories of $r(s)$ in the variable
interval $u\in [-2,2]$. Selected curves $r(s)$ for $\lambda=3,~2$,
and 1, respectively. Dots locate the current values of the
statefinder pair $\{s,r\}$. }
\end{center}
\end{figure}
%%%%%%%%%%%%%%%%%%%%%%%%%%%%%%%%%%%%%%%%%%%%%%%%%%%%%%%%%%%%%%%%%

We first apply a statefinder analysis on the exponential case. In
Fig.3, we show the time evolution of the statefinder pair
$\{r,s\}$. The plot is for variable interval $u\in [-2,2]$, and
the selected evolution trajectories of $r(s)$ correspond to
$\eta=2$ and $\lambda=3,~2$ and 1, respectively, and the other
model parameters are taken to be the same values as those used in
Fig.1. We see clearly that the distant from this model to LCDM
scenario is somewhat far. It is of interest to find that the
trajectory of $r(s)$ will form swirl before reaches the attractor,
which is quite different from other DE models (see
\cite{sahni,alam,gorini,zimdahl}). It is demonstrated again that
the statefinder can successfully characterize and differentiate
between various DE models. As complementarity for the diagnostic,
we also plot the evolution trajectories of statefinder pair
$\{r,q\}$ in Fig.4. We see that the cosmic acceleration is ensured
by sufficient strong coupling $\lambda$.

%%%%%%%%%%%%%%%%%%%%%%%%%%%%%%%%%%%%%%%%%%%%%%%%%%%%%%%%%%%%%%%%$
\vskip.8cm
\begin{figure}
\begin{center}
\leavevmode \epsfbox{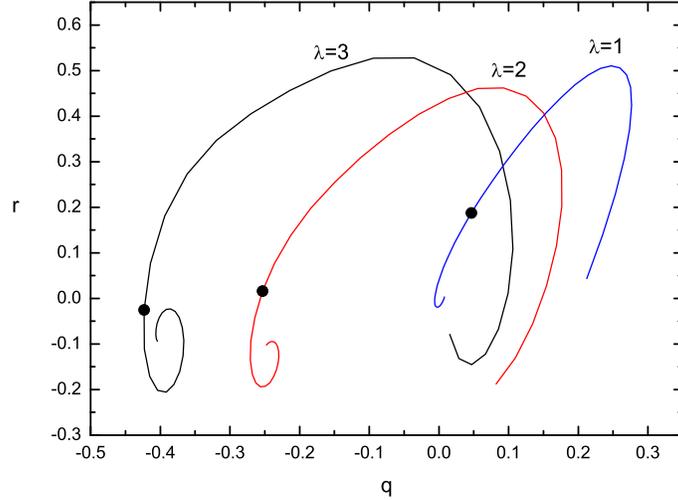} \caption[]{The $q-r$ diagram of the
exponential case: evolution trajectories of $r(q)$ in the variable
interval $u\in [-2,2]$. Selected curves $r(q)$ for $\lambda=3,~2$,
and 1, respectively. Dots locate the current values of the
statefinder pair $\{q,r\}$. }
\end{center}
\end{figure}
%%%%%%%%%%%%%%%%%%%%%%%%%%%%%%%%%%%%%%%%%%%%%%%%%%%%%%%%%%%%%%%%%

%%%%%%%%%%%%%%%%%%%%%%%%%%%%%%%%%%%%%%%%%%%%%%%%%%%%%%%%%%%%%%%%$
\vskip.8cm
\begin{figure}
\begin{center}
\leavevmode \epsfbox{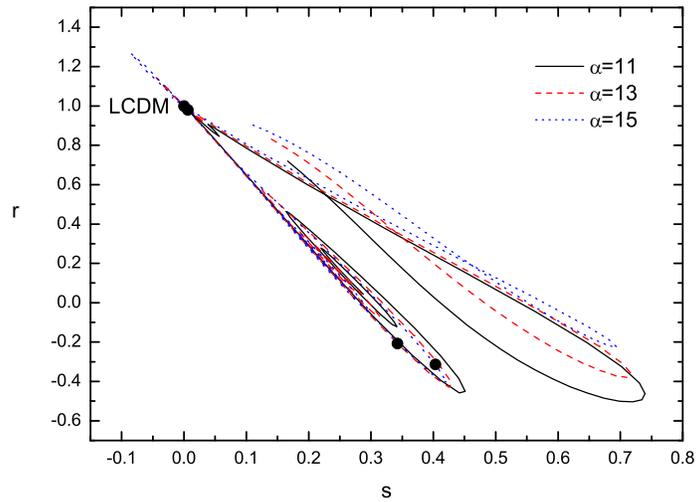} \caption[]{The $s-r$ diagram of the
power law case: evolution trajectories of $r(s)$ for the variable
interval $u\in [-2,2]$. Selected curves $r(s)$ for
$\alpha=11,~13$, and 15, respectively. Dots locate the current
values of the statefinder pair $\{s,r\}$. }
\end{center}
\end{figure}
%%%%%%%%%%%%%%%%%%%%%%%%%%%%%%%%%%%%%%%%%%%%%%%%%%%%%%%%%%%%%%%%%

Next we apply a statefinder diagnostic to the power law case. In
Fig.5, we show the time evolution of the statefinder pair
$\{r,s\}$. The plot is also for variable interval $u\in [-2,2]$,
and the selected evolution trajectories of $r(s)$ correspond to
$\beta=4$ and $\alpha=11,~13$ and 15, respectively, and the other
model parameters are as the same as those used in Fig.2. It can be
seen that the trajectories of this case will pass through LCDM
fixed point. And the swirls in this case are more evident than
those of exponential case. We also plot the evolution trajectories
of statefinder pair $\{r,q\}$ in Fig.6.

%%%%%%%%%%%%%%%%%%%%%%%%%%%%%%%%%%%%%%%%%%%%%%%%%%%%%%%%%%%%%%%%$
\vskip.8cm
\begin{figure}
\begin{center}
\leavevmode \epsfbox{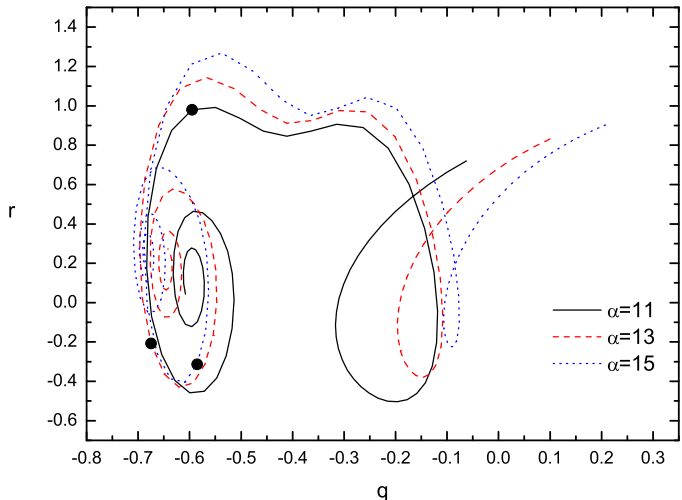} \caption[]{The $q-r$ diagram of the
power law case: evolution trajectories of $r(q)$ for the variable
interval $u\in [-2,2]$. Selected curves $r(q)$ for
$\alpha=11,~13$, and 15, respectively. Dots locate the current
values of the statefinder pair $\{q,r\}$. }
\end{center}
\end{figure}
%%%%%%%%%%%%%%%%%%%%%%%%%%%%%%%%%%%%%%%%%%%%%%%%%%%%%%%%%%%%%%%%%

In summary, we study in this Letter the statefinder of the coupled
quintessence scenario. We analyze two cases of this scenario ---
the exponential case and the power law case. It is shown that both
cases of this scenario have attractor behaviors and can thus
provide a natural solution to the cosmic coincidence problem. Then
we perform a statefinder diagnostic to both cases of this coupled
quintessence scenario. It is shown that the evolving trajectory of
this scenario in the $s-r$ plane is quite different from those of
other DE models. We hope that the future high precision
observations (e.g. SNAP) will be capable of determining these
statefinder parameters and consequently shed light on the nature
of DE.

\begin{acknowledgments}
The author would like to thank Bo Feng for useful discussions.
This work was supported by the Natural Science Foundation of China
(Grant No. 10375072).
\end{acknowledgments}

\newpage
%%%%%%%%%%%%%%%%%%%%%%%%%%%%%%%%%%%%%%%%%%%%%%%

\end{document}